\documentclass[a4paper,fleqn,10pt]{article}
\usepackage{times}
\usepackage{amssymb}
\newcommand{\text}{\mbox}
\newcommand{\notag}{\nonumber}
\begin{document}
\begin{titlepage}

    \thispagestyle{empty}
    \begin{flushright}
        \hfill{CERN-PH-TH/2012-023}\\
    \end{flushright}

    \vspace{20pt}
    \begin{center}
        { \huge{\textbf{On the Scalar Manifold of Exceptional Supergravity}}}\let\thefootnote\relax\footnotetext{Talk given at the XVII European Workshop on String Theory, held at the University of Padua, September 5-9, 2011}

        \vspace{30pt}

        {\large{\bf Sergio L. Cacciatori$^{1,4}$, Bianca L. Cerchiai$^{2,4}$, and \ Alessio Marrani$^3$}}

        \vspace{40pt}

        {$1$ \it Dipartimento di Scienze ed Alta Tecnologia,\\Universit\`a degli Studi dell'Insubria,
Via Valleggio 11, 22100 Como, Italy\\
\texttt{sergio.cacciatori@uninsubria.it}}

        \vspace{10pt}

        {$2$ \it Dipartimento di Matematica,\\
Universit\`a degli Studi di Milano,  Via Saldini 50, 20133 Milano,
Italy\\
\texttt{bianca.cerchiai@unimi.it}}

        \vspace{10pt}

        {$3$ \it Physics Department,Theory Unit, CERN, \\
        CH 1211, Geneva 23, Switzerland\\
        \texttt{alessio.marrani@cern.ch}}

\vspace{10pt}

        {$4$ \it INFN, Sezione di Milano\\
Via Celoria, 16, 20133 Milano,
Italy}

        \vspace{40pt}
\end{center}

\vspace{5pt}

\begin{abstract}
We construct two parametrizations of the non compact exceptional Lie group $G=E_{7(-25)}$, based on a fibration which has the maximal compact subgroup $K=\frac{E_6 \times U(1)}{\mathbb{Z}_3}$ as a fiber. 
It is well known that $G$ plays an important role in the $\mathcal{N}=2$ $d=4$ magic exceptional supergravity, where it describes the U-duality of the theory and where the symmetric space ${\cal M}=\frac{G}{K}$ gives the vector multiplets' scalar manifold.

First, by making use of the exponential map, we compute a realization of $\frac G K$, that is based on the $E_6$ invariant $d$-tensor, and hence exhibits the maximal possible manifest [$(E_6 \times U(1))/\mathbb{Z}_3$]-covariance. This provides a basis for the corresponding supergravity theory, which is the analogue of the Calabi-Vesentini coordinates.

Then we study the Iwasawa decomposition. Its main feature is that it is $SO(8)$-covariant and therefore it highlights the role of triality. Along the way we analyze the relevant chain of maximal embeddings which leads to $SO(8)$.

It is worth noticing that being based on the properties of a ``mixed" Freudenthal-Tits magic square, the whole procedure can be generalized to a broader class of groups of type $E_7$.
\end{abstract}

\end{titlepage}

\section{\label{sec:56}The ``mixed'' magic square and the $\mathbf{56}$ of the Lie algebra $\mathbf{\mathfrak{e}_{7(-25)}}$}

Exceptional Lie groups act as symmetries in many physical systems. In particular, non compact forms of the group $E_7$ enter as U-duality of $d=3$ and $d=4$ supergravity theories. Here we focus on the $\mathcal{N}=2$ $d=4$ magic exceptional supergravity, where the relevant real form is $G=E_{7(-25)}$. 

As the first step we need to construct the Lie algebra $\mathfrak{e}_{7(-25)}$. To this aim, we are going to follow the technique outlined in Sec. 7 of \cite{IY}, which is based on the non-symmetric ``mixed'' magic square \cite{bart-sud,gst1,tits}:

\begin{table}[h]
\caption{The ``mixed'' magic square}
\begin{center}
\vspace{-1ex}
\begin{tabular}{|c|c|c|c|c|}
\hline
& $\mathbb{R}$ & $\mathbb{C}$ & $\mathbb{H}$ & $\mathbb{O}$ \\ \hline
$\mathbb{R}$ & $SO(3)$ & $SU(3)$ & $USp(6)$ & $F_{4(-52)}$ \\ \hline
$\mathbb{C}$ & $SU(3)$ & $SU(3)\oplus SU(3)$ & $SU(6)$ & $E_{6(-78)}$ \\
\hline
$\mathbb{H}_{S}$ & $Sp(6,\mathbb{R})$ & $SU(3,3)$ & $SO^{\ast }(12)$ & $%
E_{7(-25)}$ \\ \hline
$\mathbb{O}_{S}$ & $F_{4(4)}$ & $E_{6(2)}$ & $E_{7(-5)}$ & $E_{8(-24)}$ \\
\hline
\end{tabular}
\end{center}
\label{magic}
\end{table}

\vspace{-1ex}
The rows and the columns contain the division algebras of the real numbers $\mathbb{R}$, the complex numbers $\mathbb{C}$, the quaternions $\mathbb{H}$, the octonions $\mathbb{O}$ and their split forms $\mathbb{H}_{S}$ and $\mathbb{O}_{S}$. 

Then the Tits formula gives the Lie algebra $\mathcal{L}$ corresponding to row $\mathbb{A}$ and column $\mathbb{B}$ as \cite{tits}:
\begin{equation}
\mathcal{L}\left( \mathbb{A},\mathbb{B}\right) =\text{Der}\left( \mathbb{A}\right) \oplus \text{Der}%
\left( \mathfrak{J}_{3}\left( \mathbb{B}\right) \right) \dotplus \left( \mathbb{A}^{\prime
}\otimes \mathfrak{J}_{3}^{\prime }\left( \mathbb{B}\right) \right) .
\label{Tits-formula}
\end{equation}
Here, the symbol $\oplus $ denotes direct sum of algebras, whereas $\dotplus $ stands for direct sum of vector spaces. Furthermore, Der means the linear derivations, $\mathfrak{J}_{3}\left( \mathbb{B}\right) $ denotes the rank-$3$ Jordan algebra on $\mathbb{B}$, and the priming amounts to considering only traceless elements. One of the main ingredients entering in the last term is the Lie product, which extends the multiplication to $\mathbb{A}^{\prime}\otimes \mathfrak{J}_{3}^{\prime }\left( \mathbb{B}\right)$. Its explicit expression for $\mathbb{A}=\mathbb{H}_{S}$ and $\mathbb{B}=\mathbb{O}$ can be found \textit{e.g.} in \cite{E7magic}.

For the Lie algebra of $E_{7(-25)}$ the Tits formula (\ref{Tits-formula}) yields:
\begin{equation}
\mathfrak{e}_{7\left( -25\right) }=\mathcal{L}\left( \mathbb{H}_{S},\mathbb{O}%
\right) =\text{Der}(\mathbb{H}_{S})\oplus \text{Der}(\mathfrak{J}_{3}(\mathbb{O}%
))\dotplus (\mathbb{H}_{S}^{\prime }\otimes \mathfrak{J}_{3}^{\prime }(\mathbb{O}%
))=\mathfrak{sl(2,\mathbb{R})}\oplus\mathfrak{f}_4 \dotplus \left(\mathbb{H}_{S}^{\prime } \otimes \mathfrak{J%
}_{3}^{\prime }(\mathbb{O})\right).  \label{algebra1}
\end{equation}

The second step is to identify the subalgebra $\mathfrak{K}$ generating the
maximal compact subgroup \linebreak
\mbox{$K:=(E_{6\left( -78\right) }\times U(1))/\mathbb{Z}%
_{3}$} of $E_{7(-25)}$. This can be achieved by using the Tits formula (\ref{Tits-formula}) once more to compute the manifestly $\mathfrak{f}_4$-covariant expression for $\mathfrak{e}_{6\left( -78\right)}$:
\begin{equation}
\mathfrak{e}_{6\left( -78\right) }=\mathcal{L}\left( \mathbb{C},\mathbb{O}%
\right) =\mathcal{L}\left( \mathbb{R},\mathbb{O}%
\right) \dotplus \left( i\otimes \mathfrak{J%
}_{3}^{\prime }(\mathbb{O})\right)=\text{Der}(\mathfrak{J}_{3}(\mathbb{O}))\dotplus \left( i\otimes \mathfrak{J%
}_{3}^{\prime }(\mathbb{O})\right)=\mathfrak{f}_4 \dotplus \left( i\otimes \mathfrak{J%
}_{3}^{\prime }(\mathbb{O})\right),
\label{e6-78}
\end{equation}
where we are picking the only imaginary unit $%
i\in \mathbb{H}_{S}$ which satisfies $i^{2}=-1$. Thus, we obtain:
\begin{equation}
\mathfrak{K}=ad_{i}\oplus \text{Der}(\mathfrak{J}_{3}(\mathbb{O}))\dotplus \left(
i\otimes \mathfrak{J}_{3}^{\prime }(\mathbb{O})\right) ,
\end{equation}
with $ad_{i}\in \mathbb{H}_{S}$ the \textit{adjoint action} of $i$,
generating the maximal compact subgroup $U(1)$ of the group $SL(2,\mathbb{R})$ appearing in (\ref{algebra1}). 

An explicit construction of the matrices $\phi _{I}$, $I=1,\ldots,78$, realizing the $\mathfrak{e}_{6\left( -78\right)}$ subalgebra in its irreducible representation $\mathbf{Fund}=\mathbf{27}$ has been performed e.g. in Sec. 2.1 of \cite{E6} by making use of (\ref{e6-78}) and of the explict expression  of $\frak{f}_{4(-52)}$ in its irrep. $\mathbf{Fund}=\mathbf{26}$ previously computed in \cite{F4}.

Finally, by putting together all these ingredients, we find that an explicit symplectic realization of the Lie algebra $\frak{e}_{7\left(-25\right) }$ in its irreducible representation $\mathbf{Fund}=\mathbf{56}$  is as follows \cite{e7-25}.

\noindent The generators of the maximal compact subgroup $K$ (antihermitian matrices): 
\begin{equation}
\mathfrak{e}_{6(-78)}: \quad Y_{I}=\left(
\begin{tabular}{c|c|c|c}
$\phi _{I}$ & $\overrightarrow{0}_{27}$ & $0_{27}$ & $\overrightarrow{0}%
_{27} $ \\ \hline &&&\\[-0.8em]
$\overrightarrow{0}_{27}^{T}$ & $0$ & $\overrightarrow{0}_{27}^{T}$ & $0$ \\
\hline &&&\\[-0.8em]
$0_{27}$ & $\overrightarrow{0}_{27}$ & $-\phi _{I}^{T}$ & $\overrightarrow{0}%
_{27}$ \\ \hline &&&\\[-0.8em]
$\overrightarrow{0}_{27}^{T}$ & $0$ & $\overrightarrow{0}_{27}^{T}$ & $0$
\end{tabular}
\right) , ~I=1,...,78; \label{Y_I}
\end{equation}
\begin{equation}
\mathfrak{u}(1): \quad Y_{79}=\left(
\begin{tabular}{c|c|c|c}
$\frac{i}{\sqrt{6}}I_{27}$ & $\overrightarrow{0}_{27}$ & $0_{27}$ & $%
\overrightarrow{0}_{27}$ \\ \hline &&&\\[-0.8em]
$\overrightarrow{0}_{27}^{T}$ & $-i\sqrt{\frac{3}{2}}$ & $\overrightarrow{0}%
_{27}^{T}$ & $0$ \\ \hline &&&\\[-0.8em]
$0_{27}$ & $\overrightarrow{0}_{27}$ & $-\frac{i}{\sqrt{6}}I_{27}$ & $%
\overrightarrow{0}_{27}$ \\ \hline &&&\\[-0.8em]
$\overrightarrow{0}_{27}^{T}$ & $0$ & $\overrightarrow{0}_{27}^{T}$ & $i%
\sqrt{\frac{3}{2}}$
\end{tabular}
\right) ;  \label{Y_79}
\end{equation}
The generators of the coset ${\cal M}=G/K$ (hermitian matrices): 
\begin{equation}
Y_{\alpha +79}=\frac{1}{2}\left(
\begin{tabular}{c|c|c|c}
$0_{27}$ & $\overrightarrow{0}_{27}$ & $2iA_{\alpha }$ & $i\sqrt{2}%
\overrightarrow{e}_{\alpha }$ \\ \hline
$\overrightarrow{0}_{27}^{T}$ & $0$ & $i\sqrt{2}\overrightarrow{e}_{\alpha
}^{T}$ & $0$ \\ \hline &&&\\[-0.8em]
$-2iA_{\alpha }$ & $-i\sqrt{2}\overrightarrow{e}_{\alpha }$ & $0_{27}$ & $%
\overrightarrow{0}_{27}$ \\ \hline &&&\\[-0.8em]
$-i\sqrt{2}\overrightarrow{e}_{\alpha }^{T}$ & $0$ & $\overrightarrow{0}%
_{27}^{T}$ & $0$
\end{tabular}
\right) , ~\alpha =1,...,27;  \label{Y_alpha+79}
\end{equation}
\begin{equation}
Y_{\alpha +106}=\frac{1}{2}\left(
\begin{tabular}{c|c|c|c}
$0_{27}$ & $\overrightarrow{0}_{27}$ & $-2A_{\alpha }$ & $\sqrt{2}%
\overrightarrow{e}_{\alpha }$ \\ \hline &&&\\[-0.8em]
$\overrightarrow{0}_{27}^{T}$ & $0$ & $\sqrt{2}\overrightarrow{e}_{\alpha
}^{T}$ & $0$ \\ \hline &&&\\[-0.8em]
$-2A_{\alpha }$ & $\sqrt{2}\overrightarrow{e}_{\alpha }$ & $0_{27}$ & $%
\overrightarrow{0}_{27}$ \\ \hline &&&\\[-0.8em]
$\sqrt{2}\overrightarrow{e}_{\alpha }^{T}$ & $0$ & $\overrightarrow{0}%
_{27}^{T}$ & $0$
\end{tabular}
\right) , ~\alpha =1,...,27.  \label{Y_alpha+106}
\end{equation}

Here $I_{n}$ is the $n\times n$ identity matrix, $0_{27}$ is the $27\times
27$ null matrix, $\overrightarrow{0}_{n}$ is the zero vector in $\mathbb{R}%
^{n}$, and $\overrightarrow{e}_{\alpha }$, $\alpha =1,...,27$, is the
canonical basis of $\mathbb{R}^{27}$.

The matrices $A_{\protect\alpha }$ are defined in terms of the $d$-tensor of the $\mathbf{27}$ of $E_{6\left( -78\right)}$.
There is a cubic form, which is defined for any $%
j_{1},j_{2},j_{3}\in \mathfrak{J}_3(\mathbb{O})$ as \cite{freudenthal, Small-Orbits-Phys,Small-Orbits-Maths}:
\begin{equation}
Det(j_{1},j_{2},j_{3}) :=\frac{1}{3}\text{Tr}(j_{1}\circ j_{2}\circ j_{3})
+\frac{1}{6}\text{Tr}(j_{1})\text{Tr}(j_{2})\text{Tr}(j_{3})-%
\frac{1}{6}\Big( \text{Tr}(j_{1})\text{Tr}(j_{2}\circ j_{3})+\text{cyclic perm.} \Big),
\label{determinant}
\end{equation}
where $\circ$ is the product in $\mathfrak{J}_3(\mathbb{O})$.
By choosing a basis $\{j_{a}\}_{a=1,...,26}$ of $\mathfrak{J}_{3}^{\prime }(%
\mathbb{O})$ normalized as $\langle j_{a},j_{b}\rangle :=\mbox{Tr}( j_a \circ j_b)=2\delta _{ab}$, a
completion to a basis for $\mathfrak{J}_{3}(\mathbb{O})$ can be obtained by
adding $j_{27}=\sqrt{\frac{2}{3}}I_{3}$. Then the matrices $A_{\alpha }$'s are $27\times 27 $ symmetric matrices, whose components, explicitly
computed in \cite{E7magic}, satisfy the following relation \cite{freudenthal}:
\begin{equation}
(A_{\alpha })_{\ \gamma }^{\beta }=\frac{3}{2}\,Det(j_{\alpha
},j_{\gamma },j_{\beta })=:\frac{1}{\sqrt{2}}\,d_{\alpha \gamma \beta },
\label{d-tensor}
\end{equation}
where $d_{\alpha \gamma \beta }=d_{\left( \alpha \gamma \beta \right) }$ is
the totally symmetric rank-$3$ invariant $d$-tensor of the $\mathbf{27}$ of
of $E_{6\left( -78\right) }$, with a normalization suitable to match $%
Det(j_{\alpha },j_{\gamma },j_{\beta })$ given by (\ref{determinant}).
Whenever the choice of the basis $%
\{j_{\alpha }\}$ is exploited in order to distinguish the identity matrix
from the traceless ones, the $d_{\alpha \beta \gamma }$ of $E_{6}$ has a
maximal manifestly $F_{4\left( -52\right) }$-invariance only.
However, it is crucial to point out that, being expressed only in terms of the invariant $d$-tensor, the result (\ref{d-tensor}) does not depend on the particular
choice of the basis $\{j_{\alpha }\}$. Thus, the expressions of $Y_{\alpha
+79}$ (\ref{Y_alpha+79}) and of $Y_{\alpha +106}$ (\ref{Y_alpha+106})
exhibit the maximal manifest compact $\left[(E_6 \times U(1))/\mathbb{Z}_3 \right] $-covariance.

A couple of remarks on the properties of the matrices $Y_A$'s are in order.
The first is that they satisify: 
\begin{equation}
Y_A \in \mathfrak{usp}(28,28), ~ A=1,\ldots,133.
\label{usp}
\end{equation}
Moreover, in order to guarantee that the period of the maximal torus in the $E_6$
subgroup equals $4 \pi$, the standard choice for the period of the spin representations of the orthogonal subgroups \cite{F4, E6}, the matrices $Y_{A}$'s are orthonormalized as
$\langle Y,Y^{\prime }\rangle _{\mathbf{56}}:=\displaystyle{\frac{1}{12}}\text{Tr}%
(YY^{\prime })$
with signature $(-^{79},+^{54})$. As a consequence, the components $(A_{\alpha })_{\ \gamma }^{\beta }:=A_{\alpha \beta \gamma }$ are normalized as $A_{\alpha \beta \gamma }A^{\eta \beta \gamma }=5\delta _{\alpha }^{\eta }$.

This is consistent with the normalization of the $d$-tensor (of $E_{6\left(
-26\right) }$) adopted \textit{e.g.} in \cite{adfl}, which is dictated by the expression $
f\left( z\right) :=\frac{1}{3!}d_{\alpha \beta \gamma }z^{\alpha }z^{\beta
}z^{\gamma }$ for the K\"{a}hler-invariant ($\left( X^{0}\right) ^{2}$-rescaled) holomorphic prepotential function characterizing special K\"{a}hler geometry (see \textit{e.g.} \cite{Strominger-SKG,dWVVP,N=2-Big}, and Refs. therein).

\section{\label{manifest}Manifestly $\mathbf{\left[ (E_6 \times U(1))/\mathbb{Z}_3 \right]}$-covariant Construction of the Coset $\mathbf{\mathcal{M}}$}

In this Section we construct a manifestly $\left[(E_6 \times U(1))/\mathbb{Z}_3 \right]$-covariant parametrization of the symmetric space $\mathcal{M}=\frac {E_{7(-25)}}{(E_{6(-78)}\times U(1))/\mathbb{Z}_3}$. As we have seen in the previous Sec. \ref{sec:56},
it is generated by the matrices $Y_{79+I}$, (\ref{Y_alpha+79}) and (\ref{Y_alpha+106}) with $I=1,\ldots ,54$. Through the exponential mapping, it can be defined as follows:
\begin{equation}
\mathcal{M}:=\exp \left( \sum_{\alpha =1}^{27}x_{\alpha }Y_{106+\alpha
}+y_{\alpha }Y_{79+\alpha }\right) , \mbox{ with } x_{\alpha} \in \mathbb{R}, y_{\alpha} \in \mathbb{R}, \mbox{ for } \alpha =1,\ldots, 27. \label{coset1}
\end{equation}

In order to make the complex structure of $\mathcal{M}$ manifest, it is convenient to introduce the following complex linear combinations of the matrices:
\begin{equation}
\zeta _{\alpha }:=\frac{1}{\sqrt{2}}\left( Y_{79+\alpha
}+i\;Y_{106+\alpha }\right) , \quad
\bar{\zeta}_{\alpha }:=\frac{1}{\sqrt{2}}\left( Y_{79+\alpha
}-i\;Y_{106+\alpha }\right)
\end{equation}
together with the corresponding complex linear combinations of the parameters:
\begin{equation}
z_{\alpha }:=\frac{1}{\sqrt{2}}(y_{\alpha }+i\,x_{\alpha }), \quad
\bar{z}_{\alpha }:=\frac{1}{\sqrt{2}}(y_{\alpha }-i\,x_{\alpha }),
\label{z-alpha}
\end{equation}
which allows to rewrite (\ref{coset1}) as
\begin{equation}
\mathcal{M}:=\exp \left( \sum_{\alpha =1}^{27}\bar{z}_{\alpha }\zeta
_{\alpha }+z_{\alpha }\bar{\zeta}_{\alpha }\right) .  \label{coset2}
\end{equation}
By introducing the $27$ dimensional complex vector $\displaystyle{z:={\sum_{\alpha =1}^{27}}z_{\alpha }\vec{e}_{\alpha }}$, describing the scalar fields,
and the $28\times 28$ matrix
$\mathcal{A}:=\left(
\begin{array}{c|c}
\displaystyle{-\sqrt{2}{\sum_{\alpha =1}^{27}}\bar{z}_{\alpha }A_{\alpha }} & z \\[3ex] \hline
&  \\[-1.6ex]
\;z^{T} & 0
\end{array}
\right)$,
the expression for $\mathcal{M}$ (\ref{coset2}) enjoys the simple form:
\begin{equation}
\mathcal{M}:=\exp \left(
\begin{array}{c|c}
0 & \mathcal{A} \\ \hline
&  \\[-1.6ex]
\mathcal{A}^{\dagger } & 0
\end{array}
\right) =\left(
\begin{array}{c|c}
\text{Ch}(\sqrt{\mathcal{AA}^{\dagger }}) & \displaystyle{\mathcal{A}%
\frac{\text{Sh}(\sqrt{\mathcal{A}^{\dagger }\mathcal{A}})}{\sqrt{\mathcal{A}%
^{\dagger }\mathcal{A}}}} \\[2.5ex] \hline
&  \\[-1.2ex]
\displaystyle{\frac{\text{Sh}(\sqrt{\mathcal{AA}^{\dagger }})
}{\sqrt{\mathcal{AA}^{\dagger }}} \mathcal{A}^{\dagger }} & \text{Ch}(\sqrt{\mathcal{A}%
^{\dagger }\mathcal{A}})
\end{array}
\right) .  \label{dcoset2}
\end{equation}
This is a Hermitian matrix, of the same form as the finite coset
representative worked out \cite{DeWitNicolai} for the \textit{split} (%
\textit{i.e.} maximally non-compact) counterpart $\mathcal{M}_{\mathcal{N}=8}=\frac{E_{7\left( 7\right) }}{SU\left( 8\right) /%
\mathbb{Z}_{2}},$
which is the scalar manifold of \textit{maximal} $\mathcal{N}=8$, $D=4$
supergravity, associated to $\mathfrak{J}_{3}\left( \mathbb{O}_{S}\right) $. However, while $\mathcal{M}_{\mathcal{N}=8}$ is real, because of (\ref{usp}) $\mathcal{M}$ is an element of $USp\left( 28,28\right)$. 

By using the machinery of \textit{special K\"{a}hler geometry} (see \textit{%
e.g.} \cite{Strominger-SKG,dWVVP,N=2-Big}, and Refs. therein), the
symplectic sections defining the \textit{symplectic frame} associated to the
coset parametrization introduced above can be directly read from (\ref{dcoset2}):
\begin{equation}
\mathcal{M}=:\left(
\begin{array}{c|c}
u_{i}^{\Lambda }\left( z,\overline{z}\right) & v_{i\Lambda }\left( z,%
\overline{z}\right) \\[0.8ex] \hline
&  \\[-1ex]
v^{i\Lambda }\left( z,\overline{z}\right) & u_{\Lambda }^{i}\left( z,%
\overline{z}\right)
\end{array}
\right) ,  \label{M-call-par}
\end{equation}
where the symplectic index $\Lambda =0,1,...27$ (with $0$ pertaining to the $%
\mathcal{N}=2$, $D=4$ graviphoton), and $i=\overline{\alpha },28$. Thus, the
symplectic sections read (see \textit{e.g.} \cite{CDF-rev,N=2-Big} and Refs.
therein; subscript ``$28$'' omitted):
\begin{eqnarray}
f_{i}^{\Lambda } &:&=\frac{1}{\sqrt{2}}\left( u+v\right) _{i}^{\Lambda
}=\left( \overline{f}_{\overline{\alpha }}^{\Lambda },f^{\Lambda }\right)
:=\left( \overline{\mathcal{D}}_{\overline{\alpha }}\overline{L}^{\Lambda
},L^{\Lambda }\right) =\exp \left( \frac{1}{2}K\right) \left( \overline{%
\mathcal{D}}_{\overline{\alpha }}\overline{X}^{\Lambda },X^{\Lambda }\right)
;  \label{f-sect} \\
h_{i\Lambda } &:&=-\frac{i}{\sqrt{2}}\left( u-v\right) _{i\Lambda }=\left(
\overline{h}_{\overline{\alpha }\mid \Lambda },h_{\Lambda }\right) :=\left(
\overline{\mathcal{D}}_{\overline{\alpha }}\overline{M}_{\Lambda
},M_{\Lambda }\right) =\exp \left( \frac{1}{2}K\right) \left( \overline{%
\mathcal{D}}_{\overline{\alpha }}\overline{F}_{\Lambda },F_{\Lambda }\right)
,  \label{h-sect}
\end{eqnarray}
where $\mathcal{D}$ is the K\"{a}hler-covariant differential operator,
\begin{equation}
\mathcal{V}:=\left( L^{\Lambda },M_{\Lambda }\right) ^{T}=\exp \left( \frac{1%
}{2}K\right) \left( X^{\Lambda },F_{\Lambda }\right) ^{T}  \label{V-call}
\end{equation}
is the symplectic vector of K\"{a}hler-covariantly holomorphic sections, and
\begin{equation}
K:=-\ln \left[ i\left( \overline{X}^{\Lambda }F_{\Lambda }-X^{\Lambda }%
\overline{F}_{\Lambda }\right) \right]  \label{K}
\end{equation}
is the K\"{a}hler potential determining the corresponding geometry.
A more explicit expression for (\ref{dcoset2}) would be needed in order to check that the prepotential $F$ does not exist (i.e., $2F=X^{\Lambda }F_{\Lambda }=0$ \cite{CDFVP}) in the symplectic frame we have just introduced, which can be considered the analogue of the Calabi-Vesentini basis \cite{CV,CDFVP}, whose manifest covariance is the maximal one.

\section{\label{Iwa-Exc}The Iwasawa Decomposition and the role of triality}

Now we are going to find another parametrization for the coset $\mathcal{M}$, provided by the Iwasawa decomposition. In this case the maximal manifest covariance is broken down to a subgroup $SO(8)$, thus providing a manifestly triality-symmetric description.

The manifold $\mathcal{M}$ has rank $3$, which means
that the maximal dimension of the intersection between a Cartan subalgebra
of $E_{7\left( -25\right) }$ and the generators of $\mathcal{M}$ is $3$. In particular, we can pick $3$ such generators to be the diagonal generators of
the Jordan algebra $\mathfrak{J}_{3}\left( \mathbb{O}\right) $ itself, namely $h_1=Y_{123}$, $h_2=Y_{132}$ and $h_3=Y_{133}$.

The following step is to determine a basis $\mathcal{W}_{+}$ of $54-3=51$ positive roots $\lambda _{i}^{+}$, $i=1,\ldots ,51$ with respect to $\mathfrak{H}_{3}$. Then the Iwasawa decomposition of the coset $\mathcal{M}$ is defined as:
\begin{equation}
\mathcal{M}:=\exp (x_{1}h_{1}+x_{2}h_{2}+x_{3}h_{3})\exp
(\sum_{i=1}^{51}y_{i}\lambda _{i}^{+}).  \label{Iwasawa}
\end{equation}

As anticipated, one of its main features is that since the elements $h_{1}$, $h_{2}$, $h_{3} \in \mathfrak{h}_{3}$ commute with a 28-dimensional subalgebra $\mathfrak so(8)$, the Iwasawa parametrization of $\mathcal{M}$ exhibits a maximal manifest covariance given by $SO(8)$. Therefore, the $51$-dimensional linear space $\Lambda _{+}$ generated by the positive roots $\mathcal{W}_{+}$ is invariant under the (adjoint) action of $SO(8)$, and it decomposes into irreps. of $SO(8)$ as:
\begin{equation}
\Lambda _{+}=\mathbf{1}^{3}+\mathbf{8}_{v}^{2}+\mathbf{8}_{c}^{2}+\mathbf{8}%
_{s}^{2},  \label{triality}
\end{equation}
which is a manifestly triality-symmetric decomposition. In particular, at the level of algebras $\mathfrak{so}(8)=\mathfrak{tri}(\mathbb{O})$ with the automorphism group Aut$\left( \mathbf{t} \left( \mathbb{O}\right) \right) =Spin\left( 8\right) $ of the normed triality over the octonions $\mathbb{O}$ \cite{Baez}.

It is worth remarking that the appearance of the square for the
three $\mathbf{8}$ irreps. in (\ref{triality}) is a consequence of the
complex (special K\"ahler) structure of the coset $\mathcal{M}$.

Moreover, it should be observed that the $SO\left( 8\right)$ entering in (\ref{triality}) can be identified as:
\begin{equation}
SO\left( 8\right) \subset \left[ (SO(10)\times U(1))\cap F_{4(-52)}\right] .
\label{SO(8)-def}
\end{equation}
This can be understood by noticing that it can be obtained from both the following chains of maximal symmetric embeddings \cite{Gilmore}:
\begin{equation}
E_{7\left( -25\right) }
\supset
E_{6\left( -78\right) }\times U\left(
1\right) ^{\prime } 
\supset SO\left( 10\right) \times U\left( 1\right) ^{\prime }\times
U\left( 1\right) ^{\prime \prime }
\supset
SO\left( 8\right) \times U(1)^{\prime }\times U(1)^{\prime \prime
}\times U(1)^{\prime \prime \prime } \label{embedding1}
\end{equation}
and
\begin{equation}
E_{7\left( -25\right) } \supset E_{6\left( -78\right) }\times U\left(1\right) ^{\prime } 
\supset F_{4\left( -52\right) }\times U\left( 1\right) ^{\prime }
\supset SO\left( 9\right) \times U\left( 1\right) ^{\prime } 
\supset SO\left( 8\right) \times U\left( 1\right) ^{\prime }.
\label{embedding2}
\end{equation}
In the last line of (\ref{embedding1}) the first two $U\left( 1\right) $
factors have the physical meaning of ``extra'' $T$-dualities generated by
the Kaluza-Klein reductions, respectively $D=5\rightarrow D=4$, and $%
D=6\rightarrow D=5$. 

Denoting with subscripts $U\left( 1\right) $-charges, the adjoint irrep. $\mathbf{133}$ of $E_{7\left( -25\right)
}$ branches according to (\ref{embedding1}) as (see \textit{e.g.} \cite{Slansky}):
\begin{eqnarray}
\mathbf{133} &=&\mathbf{78}_{0}+\mathbf{1}_{0}+\mathbf{27}_{-2}+\mathbf{27}%
_{+2}^{\prime }  \notag \\
&&  \notag \\
&=&\mathbf{1}_{0,0}+\mathbf{16}_{0,-3}+\mathbf{16}_{0,+3}^{\prime }+\mathbf{%
45}_{0,0}+\mathbf{1}_{0,0}  \notag \\
&&+\mathbf{1}_{-2,+4}+\mathbf{10}_{-2,-2}+\mathbf{16}_{-2,+1}  \\
&&+\mathbf{1}_{+2,-4}+\mathbf{10}_{+2,+2}+\mathbf{16}_{+2,-1}^{\prime }
\notag \\
&&  \notag \\
&=&\mathbf{1}_{0,0,0}+\mathbf{8}_{c,0,-3,1}+\mathbf{8}_{s,0,-3,-1}+\mathbf{8}%
_{c,0,+3,-1}+\mathbf{8}_{s,0,+3,+1}  \notag \\
&&+\mathbf{1}_{0,0,0}+\mathbf{8}_{v,0,0,+2}+\mathbf{8}_{v,0,0,-2}+\mathbf{28}%
_{0,0,0}+\mathbf{1}_{0,0,0}  \notag \\
&&+\mathbf{1}_{-2,+4,0}+\mathbf{1}_{-2,-2,+2}+\mathbf{1}_{-2,-2,-2}+\mathbf{8%
}_{v,-2,-2,0}+\mathbf{8}_{c,-2,+1,+1}+\mathbf{8}_{s,-2,+1,-1}  \notag \\
&&+\mathbf{1}_{+2,-4,0}+\mathbf{1}_{+2,+2,-2}+\mathbf{1}_{+2,+2,+2}+\mathbf{8%
}_{v,+2,+2,0}+\mathbf{8}_{c,+2,-1,-1}+\mathbf{8}_{s,+2,-1,+1}.\notag
\end{eqnarray}

\section{Final Remarks}

It is very interesting to remark that being based only on the algebraic properties of the ``mixed" Freudenthal-Tits magic square in Table \ref{magic}, the construction of the basis with the maximal possible covariance  (\ref{dcoset2}) and the computation of the Iwasawa decomposition (\ref{Iwasawa}) described here can be both generalized \cite{e7-25} at least to a broader class of minimally non-compact, simple groups of type $E_7$ {\cite{Brown}. 
Moreover, it also turns out that, like for $\mathcal{M}$, in all these cases the
maximal covariance (at least at the Lie algebra level) of the Iwasawa
decomposition is given by the automorphism algebra of the
corresponding normed triality \cite{Baez}.

\section*{Acknowledgements}
The work of B.L.C. has been supported in part by the European Commission under the FP7-PEOPLE-IRG-2008 Grant No. PIRG04-GA-2008-239412 “String Theory and Noncommutative Geometry” STRING.

\end{document}